\documentclass[aps,prl,groupedaddress,twocolumn,showpacs]{revtex4}
\usepackage{epsfig}
\newcommand{\rarr}{\rightarrow}

\begin{document}


\title{Estimating dissipation from single stationary trajectories.}


\author{\'Edgar Rold\'an}
\author{Juan M.R. Parrondo}
\affiliation{Dep. F\'{\i}sica At\'omica, Molecular y Nuclear and GISC.
Universidad Complutense de Madrid. 28040-Madrid, Spain}


\date{\today}

\begin{abstract}
In this Letter we show that the time reversal asymmetry of a
stationary time series provides information about the entropy
production of the physical mechanism generating the series, even if
one ignores any detail of that mechanism. We develop estimators for
the entropy production which can detect non-equilibrium processes even when there are no measurable flows in the time series.
\end{abstract}

\pacs{05.70.Ln, 05.20.-y, 05.40.-a}

\maketitle

The relationship between irreversibility and entropy production forms
the core of thermodynamics and statistical mechanics. However, it
had not been formulated quantitatively until the recent introduction
of the Kullback-Leibler distance or relative entropy in the context
of fluctuation and work theorems \cite{kpb}. The relative entropy between two
probability distributions, $p(x)$ and $q(x)$ is defined as
\begin{equation}
 D(p||q)\equiv \sum_{x}p(x)\log
\frac{p(x)}{q(x)},
\label{rel_ent}
\end{equation}
and is a measure of their distinguishability \cite{cover}. The
average entropy production associated with a process driven by an
external agent turns to be equal to the relative entropy between the
two probability distributions describing the process running forward
and backward in time \cite{kpb,njp,seifert,ciliberto,jordan}. This
relative entropy can  be thought of as the distinguishability
between the process and its time reverse, i.e., as the
irreversibility exhibited by the process. The relationship  between
entropy production and relative entropy has been derived in
different scenarios: Hamiltonian dynamics \cite{kpb,njp} and
Langevin dynamics \cite{ciliberto}, and has also been tested in
experimental situations \cite{ciliberto}.

When applied to non-equilibrium stationary states (NESS), the entropy
production per unit time reads
\begin{equation}
\frac{\langle{\dot S}\rangle}{k}=\lim_{t\to\infty}\frac{1}{t} D\left[p\left(\left\{x(\tau)\right\}_{\tau=0}^t\right)
\right|\left|p\left(\left\{x(t-\tau)\right\}_{\tau=0}^t\right)\right]
\label{def}
\end{equation}
where $k$ is the Boltzmann constant and
$p\left(\left\{x(\tau)\right\}_{\tau=0}^t\right)$ is the probability
of observing a given trajectory $\left\{x(\tau)\right\}_{\tau=0}^t$
in phase space.  Since we focus on stationary trajectories ---where
the external forcing, if any, is constant---, there is no need of
reversing the driving in the backward process. Moreover, a
sufficiently long single trajectory can provide all the necessary
statistics to compute the relative entropy in Eq.~(\ref{def}) and
consequently the entropy production rate.

Fortunately, the full information of the trajectory in the phase
space is not always necessary. Eq.~(\ref{def}) follows immediately
from the Gallavotti-Cohen theorem \cite{cohen}, by replacing the
relative entropy between trajectories with $D(p_{S}(s)||p_{S}(-s))$,
where $p_{S}(s)$ is the probability to observe an entropy production
$s$ in a time interval $[0,t]$.  In general, the relative entropy
calculated using partial information, $\left\{\tilde
x(\tau)\right\}_{\tau=0}^t$ where $\tilde x(\tau)$ is a
non-invertible function of $x(\tau)$, only provides a lower bound on
the average entropy production \cite{kpb,footprints,jordan}. For
stationary trajectories, instead of Eq.~(\ref{def}) one obtains a
lower bound, which is met if $\tilde x(\tau)$ univocally determines
the entropy production $s$.

For discrete stationary trajectories $x_{1},\dots,x_{n}$, we can
define the relative entropy of $n$-strings as
\begin{equation}
D_{n}(p_{F}||p_{B})\equiv\sum_{x_{1},\dots,x_{n}}p(x_{1},\dots,x_{n})\log
\frac{p(x_{1},\dots,x_{n})}{p(x_{n},\dots,x_{1})}
\label{Dn}
\end{equation}
Following the above arguments, we arrive at:
\begin{equation}
\frac{\langle \dot S\rangle}{k} \geq d(p_{F}||p_{B})\equiv \lim_{n\to\infty}\frac{1}{n}D_{n}(p_{F}||p_{B}).
\label{ineq}
\end{equation}

This equation reveals a striking connection between physics and the
statistics of a time series. The l.h.s. is a purely physical
quantity (it is proportional to the average dissipated energy per
step), whereas the r.h.s is a statistical magnitude depending solely
on the data $x_{1},x_{2},\dots$, but not on the physical mechanism
generating those data. Such a connection is a generalization of the
Landauer's principle relating entropy production and logical
irreversibility \cite{kpb,landauer,gasp}. Eq.~(\ref{ineq}) extends
this principle  and suggests that we can determine the entropy
production of an arbitrary NESS by computing the relative entropy of
forward and backward trajectories. We could, for instance, determine
whether a biological process is active or passive or even estimate,
or bound, the amount of consumed ATP by measuring the relative
entropy of data generated in the process.

In this Letter we explore the feasibility of such a technique by
analyzing the validity of Eq.~(\ref{ineq}) and developing estimators
of the relative entropy. Our approach is general, but we use a
discrete flashing ratchet as a case study, wherein direct comparison
between analytical and empirical values of the relative entropy and
the entropy production is possible.  There have been previous
attempts to distinguish between equilibrium  and NESS. Martin {\em
et al}  checked the fluctuation dissipation relationship in
experimental data from hair bundles of hair cells \cite{julicher},
but this approach needs two types of data: spontaneous and forced
fluctuations. Amman {\em et al} analyzed the possibility to
discriminate between equilibrium and non-equilibrium in a three
state chemical system \cite{seifert2}. Finally, Kennel introduced in
\cite{kennel} criteria based on compression algorithms to
distinguish between symmetric and asymmetric time series in the
context of chaotic signals, without any connection to dissipation.
As we show in this Letter, relative entropy provides a more general
and simpler framework for the problem of distinguishing between
equilibrium and NESS and, moreover, yields estimations and lower
bounds on the entropy production.

Two strategies have been considered  to estimate
the relative entropy between stochastic processes: the first is
based on brute-force counting of $n$-strings, obtaining empirical
estimates of $p(x_1,\dots,x_n)$, and computing $D_{n}$ using Eq.~(\ref{Dn});
the second is based on string parsing, the basic procedure of the Lempel-Ziv compression algorithm \cite{ziv}.

The first strategy is simpler and more effective for Markov chains.
Our results indicate that this is still the case for some non-Markov
process \cite{roldanparrondo}. Consequently, we will restrict
ourselves in this Letter to estimations of relative entropy from
empirical probability distributions.

If the process and its reverse are Markovian, $p(x_{1},x_{2},\dots,x_{n})=p(x_{1})p(x_{2}|x_{1})...p(x_{n-1}|x_{n})$, the relative entropy
rate $d$ defined in Eq.~(\ref{ineq}) can be expressed in terms of
the relative entropy between distributions of substrings of size 2:
\begin{equation}
d(p_{F}||p_{B})=\sum_{x_{1},x_{2}}p(x_{1},x_{2})\log\frac{p(x_{2}|x_{1})}{p(x_{1}|x_{2})}=D_{2}-D_{1}.
\label{Dmarkov}
\end{equation}
In the specific case of a trajectory and its reverse, the one-time
statistics are identical and  $D_{1}(p_{F}||p_{B})=0$. Then for
Markovian dynamics $d(p_{F}||p_{B})=D_{2}$, which can be calculated
by frequency counting if the number of states and possible
transitions is not large. In general, if one defines
\begin{equation}
 d_{k}\equiv D_{k}-D_{k-1}
\end{equation}
then $d_k\to d$ for $k\to\infty$. The limit is reached for finite
$k$ for the so-called $k$-th order Markov chains, i.e. when blocks
of size $k$, $X_{k}\equiv (x_{n},\dots,x_{n+k-1})$, are Markovian
\cite{rached}. In this case $d(p_{F}||p_{B})= d_{k+1}=d_{k+2}=\dots$.
For more general processes, we will use the following ansatz,
proposed by in Ref. \cite{grass} for Shannon entropy estimation:
\begin{equation}
d_{k} =   d_{\infty} - c\frac{\log{k}}{k^{\gamma}},
\label{ans}
\end{equation}
where $c$ and $\gamma$ are parameters that, together with $d_{\infty}$, can
be obtained by fitting the empirical values of $d_k$ vs. $k$.

\begin{figure}
\includegraphics[width=2.4cm,angle=270]{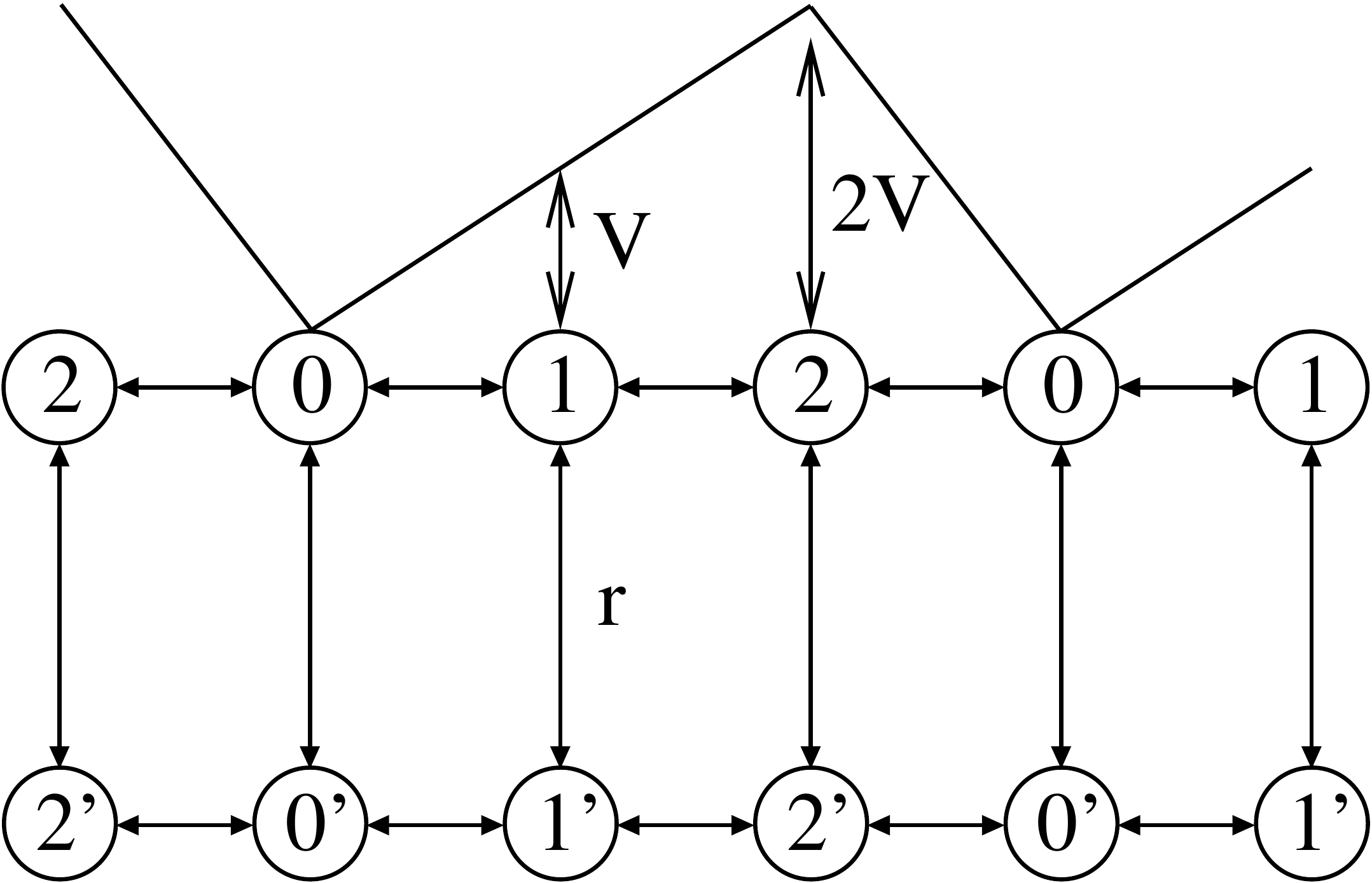}
\caption{Discrete ratchet scheme. Particles can jump between the
states $i\rightarrow j$, $i^{\prime} \rightarrow j^{\prime}$, and
$i\rightarrow i^{\prime}$ in a flashing asymmetric potential of
height $2V$ with periodic boundary conditions.  The switching rate
of the potential is $r$.} \label{dr}
\end{figure}

\begin{figure}
\includegraphics[width=6cm]{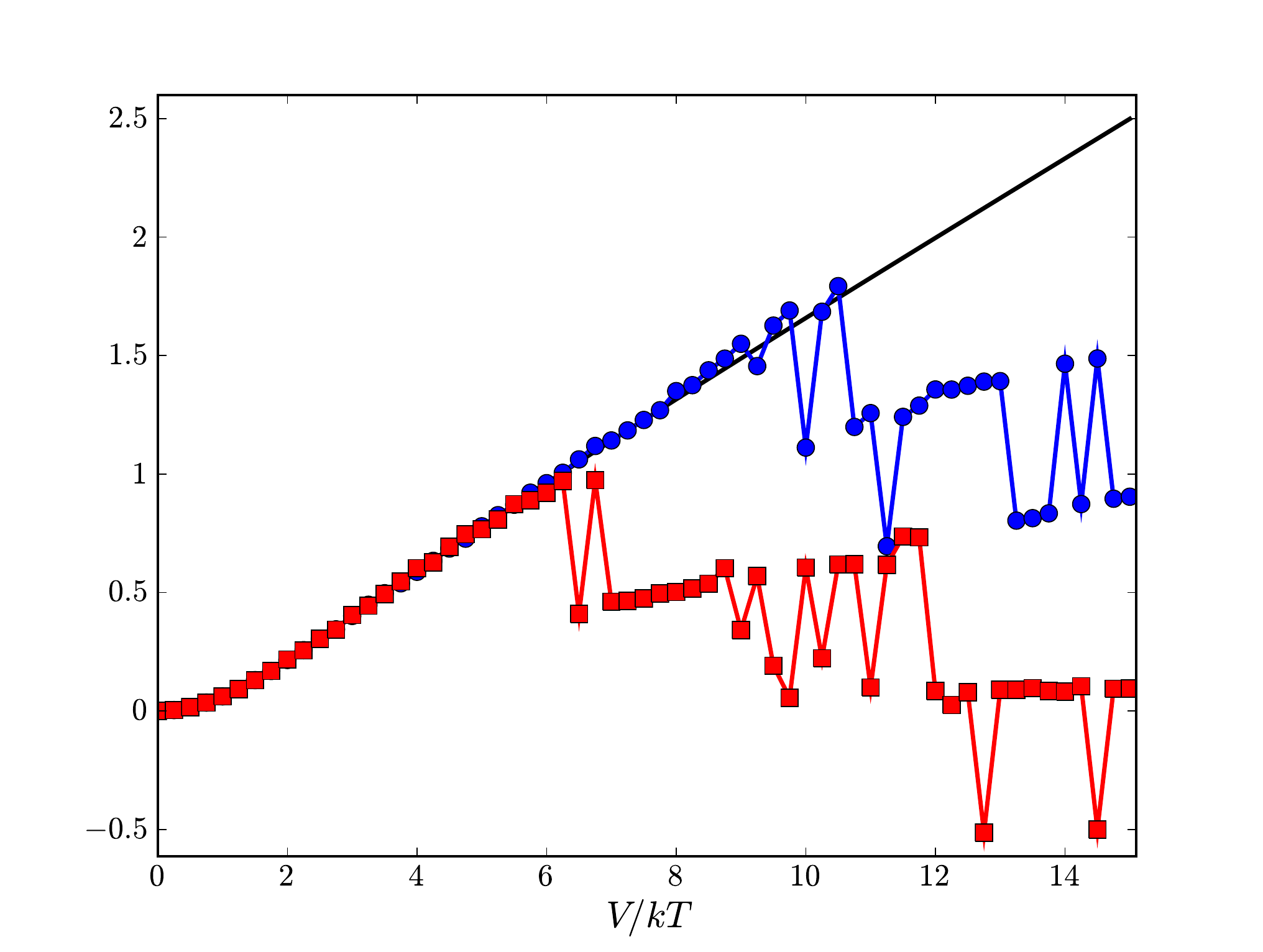}
\caption{Average dissipation per step (in units of $kT$) in the flashing ratchet ($r=1$)
and different estimations of relative entropy using a trajectory with
$n=10^{6}$ steps and full information,
as a function of  $V/kT$:
analytical calculation of the average dissipation (black line), $d_{2}$ (blue circles), $d_{3}$ (red squares).}
\label{drstate}
\end{figure}

We have tested the accuracy of these estimators and  of the bound
(\ref{ineq}) in  a specific example: a discrete flashing ratchet \cite{prost},
consisting of a particle moving in a one dimensional lattice. The
particle is at temperature $T$ and moves in a periodic and
asymmetric potential of height $2V$, which is switched on and off at
a rate $r$ (see Fig.~\ref{dr}). Trajectories are described by two
variables: the position of the particle, $x=\{0,1,2\}$, and the
state of the potential (on or off), $y=\{0,1\}$.

To define the dynamics of the particle, we start with a continuous
time description based on rates of spatial jumps and switching. We
assume that the motion in each potential obeys detailed balance: $
k_{i\rightarrow j}=e^{-\beta \frac{(V_{j}-V_{i})}{2}},$ and
$k_{i'\rightarrow j'}=1$ for $i,j=0,1,2$ with $ i\neq j $. The
system is driven out of equilibrium by imposing constant switching
rates $k_{i\rightarrow i'}=k_{i'\rightarrow i}=r$, $i=0,1,2$, which
do not obey detailed balance.

We will focus on the dissipation {\em per step}: from the continuous
trajectory $(x(t),y(t))$ we generate a series $(x_n,y_n)$ comprising
the states visited by the system. That is, we drop the information
of the times when jumps or switches occur. $(x_n,y_n)$ is a Markov
chain with transition probabilities given by $p_{\alpha\to
\gamma}={k_{\alpha\to \gamma}}/{\sum_{\gamma}k_{\alpha\to\gamma}}$,
with $\alpha,\gamma=0,1,2,0',1',2'$. Introducing these
 probabilities in Eq.~(\ref{Dmarkov}),
 $d(p_{F}||p_{B})=\beta \sum \langle V_{\alpha}-V_{\gamma} \rangle$, where the sum
 runs over transitions mediated by the thermal bath, $i\to j$,
$i'\to j'$. The relative entropy turns out to be the average
dissipation {\em per step} in units of $kT$ and we recover the main
result, Eq.~(\ref{def}) \footnote{Each transition $\alpha\to\gamma$
obeying detailed balance contributes to $d_2$ as $\beta\langle
V_\alpha-V_\gamma\rangle$ which is the average entropy increase in
the bath due to the transition. This still applies for systems in
contact with several baths at different temperatures. In our case,
the constant flashing rate $r$ can be interpreted as a transition
mediated by a bath at infinite temperature $\beta=0$, whose entropy
does not change when aborbing a finite amount of energy.}. It is
also interesting to explore the relationship between $d_2$ and the
stationary flows
$J_{\alpha\gamma}=p_{\alpha\gamma}-p_{\gamma\alpha}$ between states
$\alpha,\gamma=0,1,2,0',1',2'$. If
 $J_{\alpha\gamma}\ll p_{\alpha\gamma}$, we have:
\begin{equation}
 d_{2}\simeq \sum_{\alpha\gamma} \frac{(J_{\alpha\gamma})^{2}}{2p_{\alpha\gamma}}=\sum_{\alpha <\gamma}
  \frac{(J_{\alpha\gamma})^{2}}{p_{\alpha\gamma}}.
\label{D2J}
\end{equation}
which is a well known expression of the entropy production in
continuous Markov systems \cite{ratchet}, where $d_2=d$.

\begin{figure}
\includegraphics[width=7cm]{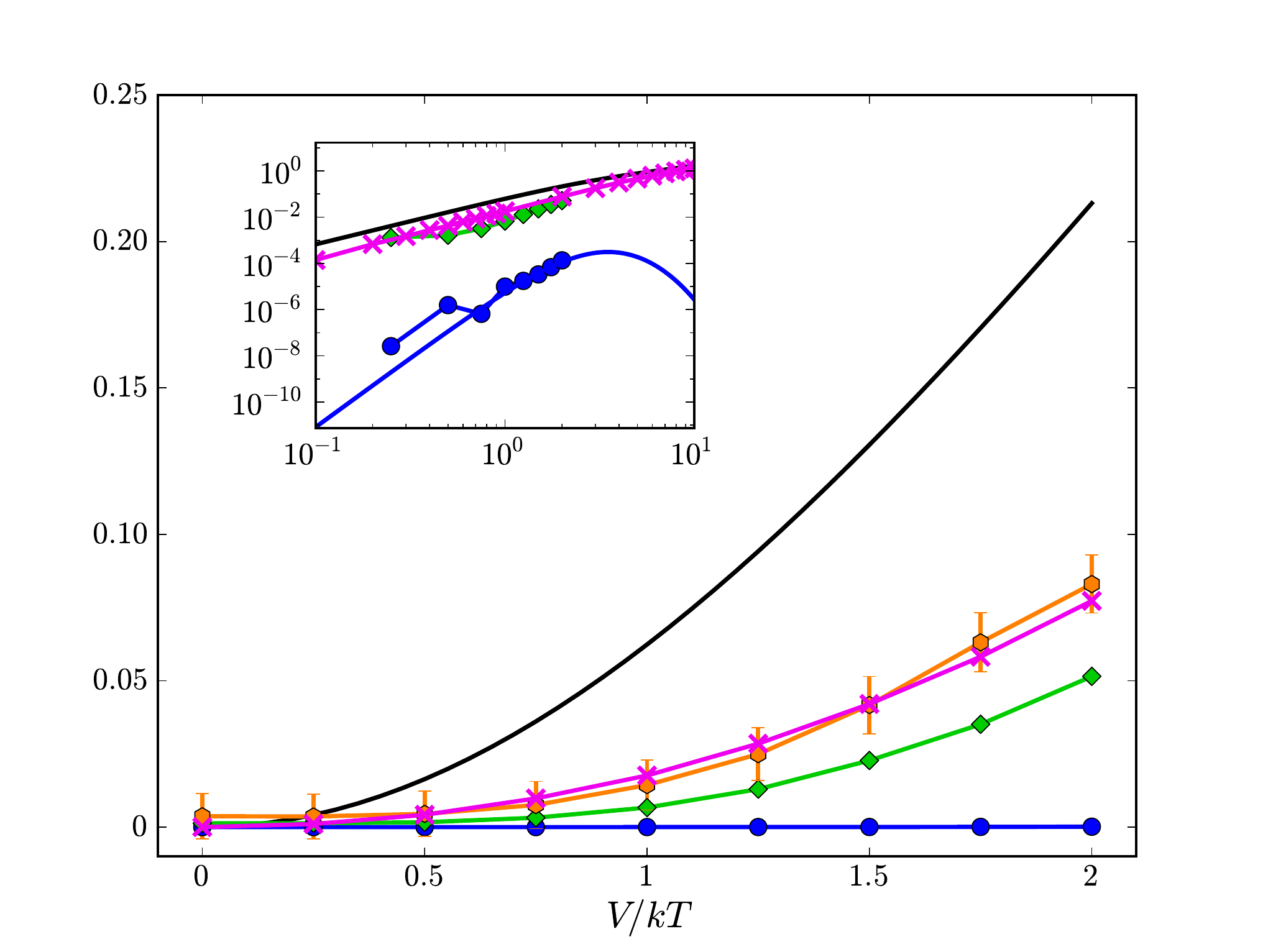}
\caption{Average dissipation per step (in units of $kT$) in the
flashing ratchet ($r=1$) and different estimations of relative
entropy using a trajectory with $n=10^{7}$ steps and partial
information (position) as a function of  $V/kT$: analytical
calculation of the average dissipation (black line),
$d_{2}$ (blue circles), $d_{9}$  (green diamonds), $d_{\infty}$
in Eq.
(\ref{ans}) (orange circles with error bars), and Monte-Carlo semi-analytical
calculation of $d$ (purple crosses). {\em Inset}. Estimators for weak potentials in a log-log plot. We have added
in the inset the analytical calculation of  $d_2$ (blue solid line).}
\label{ansfig}
\end{figure}

 Fig.~\ref{drstate}
shows the dissipation, calculated analytically by solving the
six-state Markov  chain in the stationary regime, and the
estimations discussed above. Due to Markovianity, relative
entropies, $d_k$, immediately converge $d=d_2=d_3=\dots$ and $d$ is
equal to the entropy production per step. As long as one has a good
estimation of $p(x_1,\dots,x_k)$, our approach provides accurate
values of the entropy production, which is the case for weak
potentials $V\simeq kT$. If $V\gg kT$, then uphill jumps, $0\to 1$,
$0\to 2$, and $1\to2$, are so unlikely that they do not occur in a
finite trajectory. The higher order the statistics, the earlier this
problem arises, as shown in Fig.~\ref{drstate}. The reason is that
$d_{3}$ involves probability distributions of three-step
trajectories, the sampling space is bigger and it is easier that
some transitions $i \to j \to k$ do not appear while their reverse
do. Although these jumps are very unlikely, they contribute
significantly to $d$, as shown in Fig.~\ref{drstate}, where $d_2$
and $d_3$ have been calculated by restricting the sum in $D_{k}$ to
strings  satisfying $p(x_{1}\dots x_{k})\neq 0$  and $p(x_{k}\dots
x_{1}) \neq 0$.

\begin{figure}
\includegraphics[width=7cm]{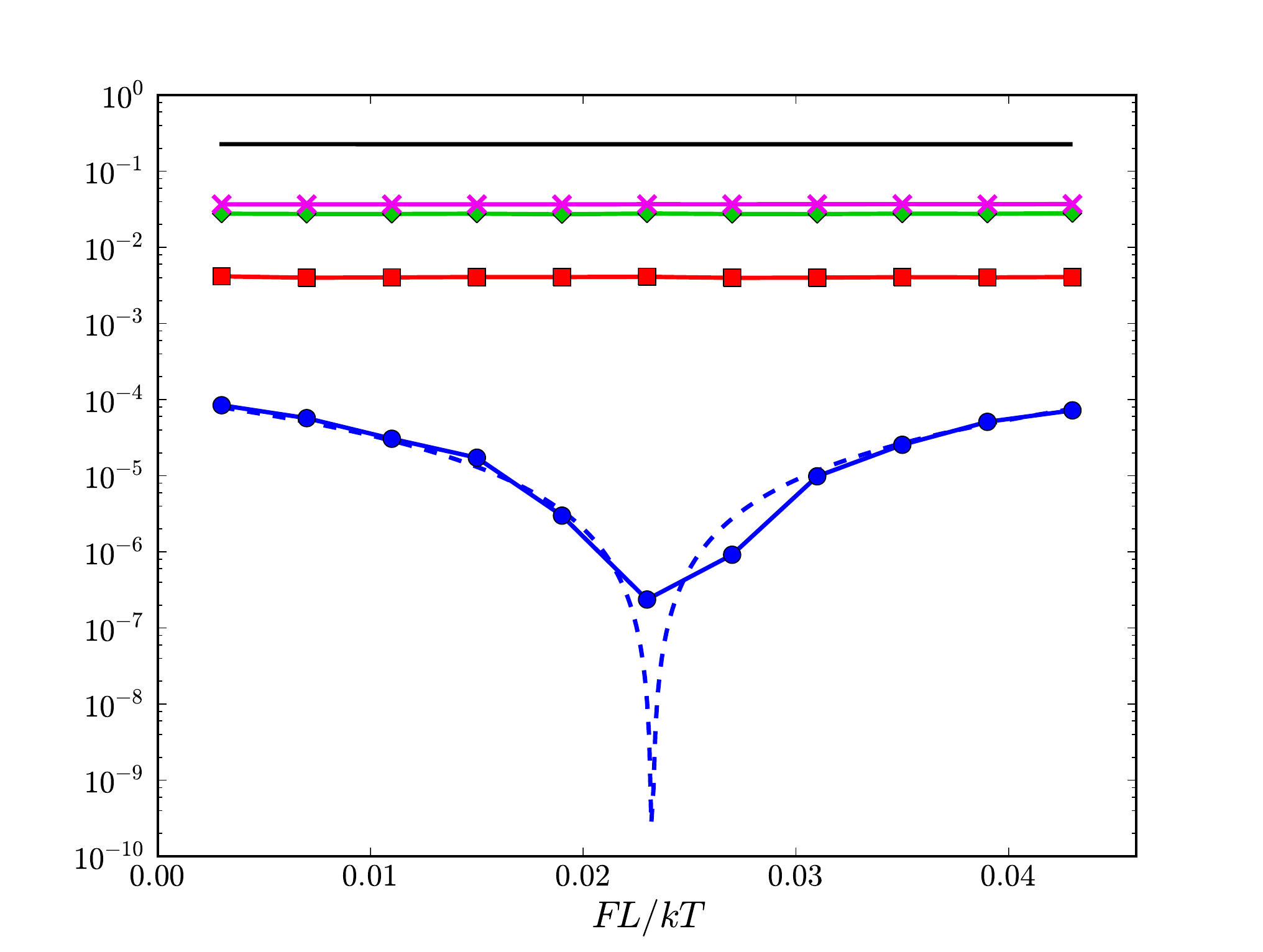}
\caption{Average dissipation per step (in units of $kT$) in the
flashing ratchet ($r=2,V=2kT$) with external force $F$ and different estimations of relative
entropy using a trajectory with $n=10^{7}$ steps  with partial
information (position): analytical
calculation of the average dissipation (black line),
$d_{2}$ (blue circles,
analytical values in blue dashed line), $d_{3}$ (red squares), $d_{9}$
(green diamonds), and  semi-analytical calculation of $d$
(purple crosses). The minimum in $d_2$ corresponds to the stall force.} \label{peak}
\end{figure}

In real applications, it is more likely that one has only partial
information of the trajectories.  To study the accuracy of our
estimators and of the inequality (\ref{ineq}) in this case, we
remove the information of the state of the potential and consider
trajectories described only by the position $\{x_{k}\}_{k=1}^{n}$,
which in general are not Markovian. As a consequence, the estimation
of the relative entropy $d(p_F||p_B)$ is more difficult, but even a
good estimation of $d$ only provides a lower bound on the relative
entropy. It is known that the Gallavotti-Cohen symmetry does not
hold in the continuous flashing ratchet if the state of the
potential is not considered \cite{lacoste}. In fact, the bound
(\ref{ineq}) can be quite loose. For instance, if $r\to \infty$,
switching is very fast and the particle moves in an effective
potential (the average of on and off) which is periodic. The
position $x_k$ becomes Markovian and the current vanishes. Using
Eq.~(\ref{D2J}) one arrives at $d=d_2=0$, whereas the dissipation
per step is non-zero.

In most cases however the bound given by Eq.~(\ref{ineq}) provides
significant information. In Fig.~\ref{ansfig} we show the estimation
of $d$ using the empirical values of $d_k$ for $k=2,9$, and the
extrapolation $d_{\infty}$ resulting from the fit of the ansatz in
Eq.~(\ref{ans}). The error bars in Fig.~\ref{ansfig} correspond to
the error in the fit with a confidence interval of 90\%. Our
estimations clearly distinguish  between the equilibrium case
($V=0$) and the NESS. The empirical $d_{k}$ with $k>3$ correctly
reproduce the order of magnitude of the actual dissipation (see
inset in Fig.~\ref{ansfig}), although they underestimate it. There
are two possible causes for this deviation: either we are
underestimating the actual relative entropy $d$, or the bound
provided by Eq.~(\ref{ineq}) is not tight. To clarify this question
we need an analytical calculation of the relative entropy between
two non-Markov processes. In our case, the relative entropy $D_{n}$
reads:
\begin{equation}
D_{n}=\left\langle \log\frac{\sum_{y_{1},\dots,y_{n}}
p(x_{1},y_{1};\dots;x_{n},y_{n})}{\sum_{y_{1},\dots,y_{n}}p(x_{n},y_{n};\dots;x_{1},y_{1})}\right\rangle
\label{dnpart}
\end{equation}
where the average is taken over all possible trajectories. The
probability distribution
$p(x_{1},y_{1};\dots;x_{n},y_{n})=p(x_{1},y_{1})\times
p(x_{2},y_{2}|x_{1},y_{1})\times \dots \times
p(x_{n},y_{n}|x_{n-1},y_{n-1})$ is known, but Eq.~(\ref{dnpart})
cannot be calculated exactly. Fortunately, the $\log$ in
Eq.~(\ref{dnpart}) is a self-averaging quantity  for large $n$
\cite{jacquet} and we can compute the average using a single long
typical trajectory \cite{roldanparrondo}. We show in
Fig.~\ref{ansfig} the value of $d$ obtained by  this Monte-Carlo
semi-analytical calculation (purple crosses), which is very close to
the estimation $d_{\infty}$ based on the ansatz Eq. (\ref{ans}).

Although the relative entropy $d$ underestimates the actual dissipation, it does reproduce
 its asymptotic behavior. Entropy production decreases
as $V^{2}$ for small $V$, so do $d_{\infty}$ and $d_9$
(see inset of Fig.~\ref{ansfig}). On the other hand, $d_{2}\propto
V^{6}$, since the current is $J\propto V^{3}$ (see Eq. (\ref{D2J})).

We have found in several instances a similar qualitative improvement
on the estimation of relative entropy when using  blocks of size
bigger than two. In particular, $d_3$ and above outperform $d_2$, which, as
indicated by Eq.~(\ref{D2J}), is equivalent to the standard
calculation of entropy production using the currents
observable from the available data; in our case, the spatial
current. For a striking illustration of this effect we add an
external force $F$ to the flashing ratchet and study dissipation and
relative entropy close to the stalling force $F_{\rm stall}$, for
which the spatial current and $d_2$ both vanish. Jumping rates
are now biased in the direction of the force, giving the following
detailed balance condition $k_{i\rarr j} / k_{j\rarr i} = e^{-\beta
(V_{j}-V_{i}-FL_{ij})}$, $L_{ij}=1$ being the distance between $i$
and $j$.

We have plotted in Fig.~\ref{peak} the real dissipation, the
analytical value of $d$ and $d_2$ and the empirical values of $d_2$,
$d_3$, and $d_9$, close to the stalling force
$F_{\rm stall}$. Recall that, for $F=F_{\rm stall}$, the position of
the particle does not exhibit any flow and its average position
remains constant. Consequently, $d_2$ or any other estimation of
entropy production based on flows will fail. However, the
relative entropy calculated using blocks of size 3 captures the non-equilibrium
 nature of the time series.

In conclusion, we have shown that the statistical properties of a
time series impose a lower bound on the entropy produced in
generating the series. This lower bound is valid even if we do not
have any access or information of the physical mechanism generating
the data. Finally, we have shown that the bound can be non-trivial,
predicting dissipation even when the data do not exhibit any
measurable flow. Our techniques could be applied to data from
different sources. In the case of biological systems, they could
help to distinguish between passive and active processes, and even
to estimate ATP consumption. On the other side, as in the case of
Landauer's principle, relative entropy can be used to ascertain the
minimal entropy production associated with a specific
behavior, such as spatiotemporal patterns, excitable systems, etc.
This in turn may influence the design of optimal devices with functionalities
 given by these behaviors.

\begin{acknowledgments}
We acknowledge financial support from Grant MOSAICO (Spanish
Government), {\em Becas de la Caixa para estudios de M\'aster en
Espa\~na}, and from the Max-Planck-Institut f\"ur Physik komplexer
Systeme  (Dresden, Germany). The main idea of the paper, namely, estimating
dissipation by relative entropy in stationary trajectories, was
suggested to us by Frank J\"ulicher and Benjamin Lindner. We also
acknowledge fruitful discussions with Luis Din\'is and Abigail Klopper.
\end{acknowledgments}

\bibliography{refs}

\end{document}